\documentstyle[12pt,iopfts]{ioplppt}
\eqnobysec
\def\rf#1{(\ref{#1})}
\def\im{\mathop{{\rm Im}}}
\def\tgam{\tilde{\gamma}}
\def\ksi#1{\stack{#1}{x}}
\def\bbox#1{{\bi #1}}
\def\myrm{\vphantom{i}}

\newcommand{\stack}[2]
 {\stackrel{\scriptstyle #1}{#2}\hspace{-3.5pt}\vphantom{#2}}

\begin{document}

\title{ Quantization of fields over de Sitter space  \\
 by the method of generalized coherent states
 }[Quantization by the generalized coherent states]
\author{S A Pol'shin\ftnote{1}{E-mail: itl593@online.kharkov.ua}}
\address{Department of Physics, Kharkov National University, \\
  Svobody Sq., 4, 61077, Kharkov, Ukraine }
\jl{1}

\date{}
\begin{abstract}
A system of generalized coherent states for the de Sitter group obeying
the Klein-Gordon equation and corresponding to the massive spin zero particles
over the de Sitter space is considered. This
 allows us to construct the quantized scalar field by the resolution over
these coherent states; the corresponding propagator is computed by the
method of analytic continuation to the complex de Sitter space and
coincides with expressions obtained previously by other methods.
Considering the case of spin~1/2 we establish the connection of the invariant
Dirac equation over the de Sitter space with irreducible representations of
the de Sitter group.  The set of solutions of this equation is
obtained in the form of the product of two different systems of generalized
coherent states for the de Sitter group.  Using  these solutions the
quantized Dirac field over de Sitter space is constructed and its propagator
is found. It is a result of action of some de Sitter invariant spinor
operator onto the spin zero propagator with an imaginary shift of a mass.
We show that the constructed
propagators possess the de Sitter-invariance and causality properties.
\end{abstract}

\section{Introduction}

In the last few years a considerable progress in the  theory of massive
scalar field over the de Sitter (dS) space  is attained due to
 using the new mathematical methods. In~\cite{24} is shown that the
two-point Wightman function ${\cal W}(x,y)$ which corresponds
to this field and obeys the conditions of causality,
dS-invariance and positive definiteness can be obtained as a
boundary value of the holomorphic function $W(z_1 ,z_2)$ defined over the
complex dS space.  In turn, the function $W(z_1 ,z_2)$ can be
represented as an integral over so-called "plane waves"; these
"plane waves" obey the
Klein-Gordon equation over the dS space and generalize the usual plane waves
over the Minkowski space.  In~\cite{dS-PLB} to examine the
quantum fields over the dS space we applied the method of generalized
coherent states (CS) which has been fruitfully used in various physical
problems (see~\cite{coher1/4} and references therein). In the mentioned paper
we showed that the above "plane waves" are CS for the dS space to within  the
coordinate-independent multiplier, and their scalar product coincides with
the two-point function considered in~\cite{24}.

Nevertheless, even in the spin zero case some questions remain open. Can we
construct the quantized field  by the expansion over the
mentioned "plane waves" in such a way that its propagator will be equal to
${\cal W}(x,y)-{\cal W}(y,x)$? What is the explicit form of this propagator?
Passing to the case of spin~1/2 we see that
 usual methods are insufficient at all
for the  consistent construction of a theory of quantized field
over the dS space. Indeed, a lot of papers were concerned with
obtaining the solutions of covariant (\cite{16/65} and references therein)
and group theoretical~\cite{87} Dirac equation over the dS space by the
method of separation of variables. However, all these solutions have a
complicated form which considerably troubles the construction of the theory
of quantized field.  Only in the little-known paper~\cite{80} the summation
over one of such a set of solutions was performed; the resulting
propagator is not dS-invariant and does not obey the causality principle. On
the other hand, in~\cite{61} a spinor propagator was found starting from the
demands of dS-invariant Dirac equation satisfaction, dS-invariance and the
boundary conditions; but the quantized field  corresponding to it was not
found; on the contrary, in the Anti-de Sitter space the quantized spinor
fields with an invariant and causal propagator was constructed long
ago~\cite{36}.

In the present paper we show that all these problems may be solved using
the method of generalized coherent states, and build  the theory of massive
quantized scalar and spinor fields over the dS space using this method.
The present paper is composed as follows. In section 2, bearing in mind the
application to the spinor field,  we
give the method of construction of CS in the maximally general form for which
the Perelomov's definition is a special case.

In section 3 we consider the  scalar field. In subsection 3.1 we
consider the dS space, its symmetry group and the classification of its
irreducible representations. The realization of the dS group as a group of
transformations of Lemaitre coordinate system is given too.
Following~\cite{dS-PLB}, in
subsection 3.2 we realize the dS group as a group of
transformations of  functions over ${\Bbb R}^3$, and then construct the CS
system for the dS space which corresponds to the massive spin zero particles
and obeys the dS-invariant Klein-Gordon equation. The scalar product of two CS
is the two-point function considered in~\cite{24}; from here its
dS-invariance follows immediately that is proved in~\cite{24} by  other
reasons.  The integral  defining this two-point function may be
regularized passing to the complex dS space. For the sake of
completeness we reproduce some results of the paper by J.Bros and
U.Moschella~\cite{24} and compute the two-point function over the
complex dS space in the explicit form. In subsection 3.3 we construct
the quantized scalar field by the expansion over CS constructed in subsection
3.2. The propagator of this field is the difference of two-point function and
the permuted one. We show that the boundary value on the real dS space  of
the two-point function computed in subsection 3.2 coincides with the Green's
function obtained previously by other methods~\cite{76,33}.  The propagator
which is the difference of two two-point functions coincides with that
obtained previously starting from the demands of the dS-invariance and the
satisfaction of the Klein-Gordon equation and the boundary
conditions~\cite{77/81/82}. Thus, the relation between  different expressions
for the propagator available in the literature is established (the review of
papers concerning the propagators over the dS space see in~\cite{caus4/5}).

In section 4 we consider the  spinor field.
In subsection 4.1 we consider the dS-invariant Dirac
equation and show that the corresponding representation of the dS group is
irreducible and falls under the classification listed in subsection 3.1.
Also we show that this equation admits the reduction to the covariant
form by the well simpler way than that proposed previously~\cite{61,gath}. In
subsection 4.2 we construct the CS system for the four-spinor representation
of the dS group in the form of $4\times 2$-matrices. Solutions of the
dS-invariant Dirac equation are the products of these CS and scalar CS
obtained in subsection 3.2. In fact, these solutions are the more
compact form of the spinor "plane waves" obtained in~\cite{dS-PLB}. The
invariance properties of these solutions allow us to construct of them a
dS-invariant two-point function and compute it passing to the complex dS
space. In subsection 4.3 we construct the quantized spinor field using these
 solutions; its propagator is expressed  by the boundary values of two-point
function obtained in subsection 4.2 and coincides with the expression
obtained {\it a priori} in~\cite{61} to within the constant multiplier.

In section 5 we briefly summarize the results of this paper.

\section{Definition of the CS system}

Let $\cal G$ is a Lie group and
${\cal G}\ni g\mapsto T(g)$ is its representation in a linear
vector space $H$ with operators $T(g)$. Consider some vector
$|\psi_0 \rangle\in H$ yielding the set of vectors
\[ \{ |\psi_g \rangle\equiv T(g)\psi_0,\ \forall g\in {\cal G} \}.\]
Define the equivalence relation $\sim$ between the vectors of the $H$
space coordinated with the  product over $H$ in the following way.
Let $|\xi'\rangle$ and $|\xi''\rangle$ are the vectors of $H$.
Then we assume the existence of a  product (which, in general, is not
the mapping of $H\times H$ to ${\Bbb C}$) such as
\[\label{defsim}
|\xi'\rangle\sim|\xi''\rangle
\Rightarrow \langle\xi'|\xi'\rangle =\langle\xi''|\xi''\rangle.\]
Consider the subgroup $\cal H$ of $\cal G$ which remains in the
rest the equivalency class generated by $|\psi_0\rangle$:
\[ h\in {\cal H}\Longleftrightarrow T(h)|\psi_0\rangle \sim
|\psi_0\rangle .\]
It is obvious that the number of unequivalent elements of the
above mentioned set
$|\psi_g \rangle$ is less than the number of elements of the group $\cal G$
because  the elements $g$ and $gh,\ h\in {\cal H}$ generate the equivalent
vectors. Then, in fact, the set of unequivalent vectors is determined by the
set of  right equivalency classes $g\cal H$ which compose the symmetric
space ${\cal G}/{\cal H}$.

The mapping ${\cal G}/{\cal H}\ni \xi \mapsto g_\xi \in {\cal G}$
such that for an arbitrary $g_1 \in {\cal G}$ the equality
\begin{equation}\label{foliation}
g_1 g_\xi=g_{\xi '}h \qquad  h\in {\cal H}
\qquad  \xi ' =\xi_{g_1}
\end{equation}
is valid, is called the lifting from the
${\cal G}/{\cal H}$  space to the $\cal G$ group,
where $\xi\mapsto \xi_g$ is the action of $\cal G$ over the
${\cal G}/{\cal H}$ space. We shall use the following simple
method of construction of liftings. Let
$\xi_\circ$ is a "standard" point of the ${\cal G}/{\cal H}$ space.
Let us denote
as $g_\xi$ the set of transformations parametrized by  points $\xi$ of the
${\cal G}/{\cal H}$ space so as $\xi=(\xi_\circ )_{g_\xi}.$ It is easily seen
that
$\xi \mapsto g_\xi$ is a lifting. Indeed, let $g_1
\in {\cal G}$ is an arbitrary transformation from the group $\cal G$. Then
the transformations
$g_1 g_\xi$ and $g_{\xi'}$ both transform the point
$\xi_\circ$ into the point $\xi'$; then the transformation
$(g_1 g_{\xi})^{-1} g_{\xi'}$ remains the point $\xi_\circ$ in the rest and
 therefore belongs to $\cal H$.

The choice of lifting is the choice of the {\it representative}
$g_\xi \in\cal G$ for each equivalency class $\xi$. Then the set of all
 unequivalent vectors $|\psi_g\rangle$ is given by the {\it coherent states}
 system
\[ |\xi\rangle =T(g_\xi) |\psi_0\rangle .\]
The major property of the CS system is its $\cal G$-invariance which follows
 from~(\ref{foliation}):
\begin{equation}\label{lor7}
T(g)|\xi\rangle \sim |\xi_g\rangle \qquad  g\in {\cal G}.
\end{equation}
The Perelomov's definition for the CS system is narrower than ours as
he suppose that $\sim$ is the equality to within the phase:
\[ |\xi'\rangle\sim|\xi''\rangle \Leftrightarrow |\xi'\rangle=\e^{\i\alpha}
|\xi''\rangle  \qquad \alpha\in {\Bbb R}.\]
In a certain sense, our definition is the further generalization of
a so-called vector-like CS~\cite{coher10}.  Another difference of our
definition from  Perelomov's one is that we, following~\cite{coher10}, do not
assume the compactness of the $\cal H$ subgroup.

\section{Scalar field}

\subsection{Representations of the de Sitter group}

The dS space is a four-dimensional hyperboloid determined by the equation
$\eta _{AB}x^{A}x^{B}=-R^{2}$
in the five-dimensional space
with the pseudo-euclidean metric
$\eta_{AB} \quad (A, B, \ldots=0\ldots 3, 5)$ of signature $(+ - - - -)$.
Except the explicitly covariant vierbein indices, all the ones are raised and
lowered by the Galilean metric tensors $\eta_{AB}$ and $\eta_{\mu \nu}$.
The metric in  coordinates $x^\mu$ has the form
\begin{equation}\label{5.  21}
 g_{\mu \nu}=\eta _{\mu \nu}-
\frac{x^{\mu}x^{\nu}}{R^{2}\chi ^{2}}   \qquad
g^{\mu \nu}=\eta ^{\mu \nu}+\frac{x^{\mu}x^{\nu}}{R^{2}}
\end{equation}
where $\chi  =(1+x\cdot x/R^{2})^{1/2}.$
The symmetry group of
the dS space is the dS group $SO(4,1)$ with ten generators
$J^{AB}=-J^{BA}$ obey commutation relations
\begin{equation}\label{5.  22}
[J_{AB}, J_{CD}]=\eta _{AD}J_{BC}+\eta _{BC}J_{AD}-
\eta _{AC}J_{BD}-\eta _{BD}J_{AC}.
\end{equation}
Let us denote $P^\mu =R^{-1}J^{5\mu}$; these generators correspond to
translations.

We denote the action of the arbitrary element $g\in {SO(4,1)}$ of the dS
group over the dS space  as $x\mapsto x_g$. The stationary subgroup of an
arbitrary point of dS space is $SO(3,1)$; then we can identify the dS space
with the set of equivalency classes ${SO(4,1)}/{SO(3,1)}$.

Let us construct the operators
\begin{equation}\label{3.  2}
\Pi^\pm_i=P_{i}\pm\frac{1}{R}J_{0i}.
\end{equation}
Using the commutation relations~(\ref{5. 22}) it is easy to show that
\begin{equation}\label{3.  3}
[\Pi^+_i ,\Pi^+_k ] =[\Pi^-_i ,\Pi^-_k ] =0.
\end{equation}
We can take the operators $\bPi^{+},\bPi^{-},P^{0}$ and $J_{ik}$
as a new set of generators of the dS group; they generate subgroups which we
denote as ${ \cal T}^{+}, {\cal T}^{-},{\cal T}^{0}$ and ${\cal
R}=SO(3)$, respectively. The groups ${\cal T}^{\pm}$ are abelian by the virtue
of~(\ref{3.  3}). Besides~(\ref{3. 3}), the commutation relations are
\begin{equation} \label{3.  4}
\eqalign{
{[}\Pi^+_i ,\Pi^-_k {]}=
-\frac{2}{R}P^{0}\delta_{ik}+\frac{2}{R^2}J_{ik} \qquad
{[}P^{0},J_{ik}{]}=0 \\
{[}\Pi^\pm_i ,J_{kl}{]}=
\Pi^\pm_k \delta_{il}-\Pi^\pm_l \delta_{ik} \qquad
{[}P^{0},\bPi^{\pm} {]} =\pm \frac{1}{R}\bPi^{\pm}.
}
\end{equation}

The dS group has two independent Casimir operators:
 \begin{equation}\label{5.  26}\label{5.  27}
 C_{2}=-\frac{1}{2R^{2}}J_{AB}J^{AB} \qquad
C_{4}=W_{A}W^{A}
\end{equation}
where
\begin{equation}\label{5.  28}
 W_{A}=\frac{1}{8R}\varepsilon _{ABCDE}J^{BC}J^{DE}
\end{equation}
is an analog of the Pauly--Lubanski pseudovector in the Poincar\'e group.
There are two series of the dS group irreducible representations~\cite{4/5}:

 1)$ \bpi_{p, q}$, $p=1/2, 1, 3/2, \ldots
;q=p, p-1, \ldots, 1$ and
$1/2$.  The eigenvalues of Casimir operators in this series are
\begin{equation}\label{5.  29}
\eqalign{
R^{2}C_{2}=p(p+1)+q(q-1)-2 \\
\label{5.  30}
 R^{2}C_{4}=p(p+1)q(q-1).
}
 \end{equation}

 2)$\bnu_{m, s}$.
The quantity $s$ is a spin,
 $s=0, 1/2, 1, \ldots$;the quantity $m$ corresponds to a mass, at the
integer spin $m^{2}>0$; at the half-integer spin
$m^{2}>1/4R^{2}$; at $s=0 \quad m^{2}>-2/R^{2}$.
\begin{eqnarray}\label{5.  31}
 C_{2}=-m^{2}+R^{-2}(s(s+1)-2) \\
\label{5.  32}
C_{4}=-m^{2}s(s+1).
\end{eqnarray}

The generators of five-dimensional rotations are
\[ \label{5.  33}
J^{(l)AB}=(x^{A}\eta^{BC}-x^{B}\eta^{AC})\partial_{C}.\]
As the fifth coordinate is not independent:
$x^{5}=R \chi, $ then $\partial_{5}=0$ and we obtain the generators
of scalar representation:
\begin{equation}\label{5.  34}\label{3. 4a}
P^{(l)}_{\mu}=\chi \partial _{\mu} \qquad
J^{(l)\mu \nu}=(x^{\mu}\eta ^{\nu \sigma}-
x^{\nu}\eta ^{\mu \sigma})\partial _{\sigma}.
\end{equation}
They compose the representation $\bnu_{m,0}$ since
\begin{equation}\label{5.  35}
W_{A}^{(l)}=0 \Rightarrow  C_{4}^{(l)}=0.
\end{equation}
As $(-g)^{1/2}=1/\chi,$ then for the second order Casimir operator in the
scalar representation we obtain from
~\rf{5.  26} and~\rf{5.  34}:
\[ C_{2}^{(l)}=\Box \equiv  (-g)^{-1/2}\partial
_{\mu}((-g)^{1/2}g^{\mu \nu} \partial _{\nu}).\]
 Then using~(\ref{5. 31}) we obtain
that in the representation $\bnu_{m, 0}$ the Klein-Gordon equation
\begin{equation}\label{5.  36} \label{5. 37}
(\Box +m^{2}+2R^{-2})\psi  =0
\end{equation}
is satisfied.

By the virtue of~(\ref{3.  3}) in the scalar representation the
generators~(\ref{3.  2}) are the
derivatives along certain new coordinates
called the Lemaitre coordinates:
\begin{equation}\label{3.  4b}
\bPi^\pm=\frac{\partial}{\partial {\bi y}_\pm}.
\end{equation}
Substituting~(\ref{3. 4a}) and~(\ref{3. 4b}) into r.h.s. and l.h.s
of equation (\ref{3. 2}) respectively, we obtain the connection of
${\bi y}_\pm$ with $x^\mu$. We denote a new time coordinate independent on
${\bi y}_{\pm}$  as
$y^{0}_{\pm}=\tau_{\pm}$; then the transformation rules from the old
coordinates to the new ones are
\begin{equation}\label{3.  6a}
{\bi y}_{\pm}={\bi x}\e^{\mp\tau_{\pm}/R} \qquad
\e^{\pm\tau_{\pm}/R}=\chi \pm \frac{x^0}{R}.
\end{equation}
The operator $P_{0}$   in the new coordinates takes the form
\[ P_{0}^{(l)}=\frac{\partial}{\partial \tau_{\pm}}\mp
\frac{1}{R}{\bi y}_{\pm}\frac{\partial}{\partial {\bi y}_{\pm}}.\]
The finite transformations  belonging to the subgroups
${\cal T}^{\pm}$ and ${\cal T}^{0}$,
which we denote as $\Theta_{\pm}$ and $\Theta_{0}$ respectively,
act in the scalar
representation by the following way:
\begin{equation}\label{3.  9}
\label{3.  10}
\eqalign{
 g= \Theta_{\pm}({\bi a})\equiv \exp (\bPi^{\pm}{\bi a}R) \ :
\quad \left\{
\begin{array}{l}
{{\bi y}}_{\pm}\longmapsto {{{\bi y}}'}_{\pm}={{\bi y}}_{\pm}+{{\bi a}}R \\
\tau_{\pm}\longmapsto {\tau'}_{\pm}=\tau_{\pm}
\end{array}   \right. \\
 g= \Theta_{0}(\varepsilon )\equiv \exp (P_0 \varepsilon R)\ :
\quad\left\{
\begin{array}{l}
{{{\bi y}}'}_{\pm}={{\bi y}}_{\pm}\e^{\mp \varepsilon} \\
{\tau'}_{\pm}=\tau_{\pm}+\varepsilon R.
\end{array}  \right.
}
\end{equation}
We assume that the transformations act in the order from the right to the
left.

\subsection{Scalar coherent states}
The dS group is isomorphic to the group of conformal transformations of the
three-dimensional real space.
We denote the vector of this space  as ${\bi w}$. There exist two different
conformal realizations of the dS group; the first one corresponds
to the upper sign, and the second one to the lower one in the following
formulas. The generators take the form
\begin{eqnarray}
R\bPi^{\mp}=-\frac{\partial}{\partial {{\bi w}}} \qquad
R\bPi^{\pm}={w}^{2}\frac{\partial}{\partial {{\bi w}}}-2{{\bi w}}
\left( {{\bi w}}\frac{\partial}{\partial {{\bi w}}}\right) \nonumber\\
R P_{0}=\pm {{\bi w}}\frac{\partial}{\partial {{\bi w}}} \qquad
J_{ik}={w}_{k}\frac{\partial}{\partial  {w}_{i}}-
{w}_{i}\frac{\partial}{\partial {w}_{k}}. \nonumber
\end{eqnarray}
They obey commutation relations~(\ref{3.  3}) and~(\ref{3.  4}).
 Finite transformations have the form
\begin{eqnarray}
 g=\Theta_{\mp}({{\bi a}}):\ {{\bi w}}_{g}={{\bi w}}-{{\bi a}}\nonumber \\
\label{3.  15}
g=\Theta_{\pm}({{\bi a}}):\ {{\bi w}}_{g}=
\frac{{{\bi w}}+{{\bi a}}{w}^{2}}{1+2{\bi w}{\bi a}+{w}^{2}{a}^{2}} \\
g=\Theta_{0}(\varepsilon ):\ {{\bi w}}_{g}={{\bi w}}\e^{\pm \varepsilon}.
\nonumber
\end{eqnarray}
Let us define two different representations of the dS group acting over the
space of functions dependent on ${\bi w}$:
\[\label{3.  20}
T^{\pm}_\sigma (g)f({{\bi w}})=\left(\alpha^{\pm}_{{\bi w}}(g)\right)^\sigma
f({{\bi w}}_{g^{-1}})\]
where $\sigma\in {\Bbb C}$ and
\[\alpha^{\pm}_{{\bi w}}(g)=
\det \left( \frac{\partial w^{i}_{g^{-1}}}{\partial w^{k}}
\right)^{-1/3}=
\left\{
\begin{array}{ll}
1 & g\in {\cal T}^{\mp}\circledS {\cal R} \\
\e^{\pm \varepsilon} & g=\Theta_{0}(\varepsilon) \\
1-2{\bi a}{\bi w}+a^{2} w^{2} & g=\Theta_{\pm}({{\bi a}}).
\end{array}
\right.\]
We denote these representations as $T_\sigma^\pm$. It is easily seen that the
generators in these representations are
\begin{equation}\label{3. 20a}
\eqalign{
R\bPi^{\mp}=-\frac{\partial}{\partial {{\bi w}}} \qquad
R\bPi^{\pm}={w}^{2}\frac{\partial}{\partial {{\bi w}}}-2{{\bi w}}
({{\bi w}}\frac{\partial}{\partial {{\bi w}}})+2\sigma {{\bi w}} \\
R P_{0}=\pm\left( {{\bi w}}\frac{\partial}{\partial {{\bi w}}}-\sigma \right)
\qquad
J_{ik}={w}_{k}\frac{\partial}{\partial  {w}_{i}}-
{w}_{i}\frac{\partial}{\partial {w}_{k}}.
}
\end{equation}
We define the scalar product in the space of  representation
$T_\sigma^\pm$ as follows:
\[\langle f_{1} |f_{2}\rangle =\int_{{\Bbb R}^3} \d^{3}{{\bi w}} \,
f_{1}^{*}({{\bi w}}) f_{2}({{\bi w}}).\]
It is not difficult to show that it is dS-invariant at
\[\sigma=\sigma_{0}\equiv -\frac{3}{2}-\i\mu R \qquad  \mu\in {\Bbb R}.\]
Then the representation $T^\pm_{\sigma_0}$ is unitary; but it is reducible
since we do not assume the square integrability of functions carrying
it and therefore the space contains the invariant subspace of square
integrable functions. Such an extension of the representation space is
necessary for the construction of CS with noncompact stability
subgroups~\cite{coher10}.

The equality
\[\label{3. 18a}
g_{y_{\pm}}=\Theta_{\pm}({{\bi y}}_{\pm}/R)
\Theta_{0}(\tau_{\pm}/R)\]
defines the lifting in the sense of~(\ref{foliation}) since the transformation
$g_{y_\pm}$ transforms the origin into the point with coordinates $y_\pm$.
As an equivalency relation we can take the equality. Then the vector
$|\psi_{0}\rangle$ being a Lorentz-invariant under the action of the
representation $T^\pm_\sigma$ is $|\psi_{0}\rangle
=(1-{w}^{2})^{\sigma}.$ Then we can construct the CS system
\[\label{3. 24}
|x ,\pm;\sigma\rangle =T^{\pm}_\sigma (g_{y_{\pm}(x)})|\psi_{0}\rangle \]
where we assume that  the Lemaitre coordinates are dependent on $x^\mu$
by the transformations~(\ref{3.  6a}).  The explicit form of CS as a function
of ${\bi w}$ is
\[|x ,\pm;\sigma\rangle \equiv \Phi^{(0)\pm}_{{\bi w}}(x;\sigma) =
(1-{w}^{2})^{\sigma}\varphi_{k_{{\bi w}}}^{(0)\pm}(x;\sigma)\]
where
\[\varphi_{k}^{(0)\pm}(x;\sigma)=
\left( \chi\pm \frac{k\cdot x}{R} \right)^{\sigma}\label{5. 58} \qquad
k^{\mu}_{{\bi w}}=\left( \frac{1+{w}^{2}}{1-{w}^{2}},
 \pm \frac{2{{\bi w}}}{1-{w}^{2}}\right) \]
then $k_{{\bi w}}\cdot k_{{\bi w}}=1$.
From~(\ref{lor7}) the transformation rules
\begin{equation}\label{transf}
\Phi_{{\bi w}}^{(0)\pm}(x_g ;\sigma)=\left(
\alpha^\pm_{{{\bi w}}}(g)\right)^\sigma \Phi_{{{\bi w}}'}^{(0)\pm}(x;\sigma)
\qquad  {{\bi w}}'={{\bi w}}_{g^{-1}}
\end{equation}
follow.

The functions $\varphi_{k}^{(0)\pm}(x;\sigma_0)$ obey the dS-invariant
Klein-Gordon equation~(\ref{5. 36}) and were known previously in this
capacity~\cite{24,coher9/78}.  Under $R \rightarrow \infty$ these functions
pass into the usual plane waves over the Minkowski space.

Let us consider the scalar product of two CS in the representation
$T^\pm_{\sigma_0}$; it is easily seen that the scalar products in
the representations $T^+_{\sigma_0}$ and
$T^-_{\sigma_0}$ are equal to each other. This may be proved
considering the inversion $\bbox{w}\mapsto
-\bbox{w}/{\myrm w}^2$  which yields
\[\Phi^{(0)\pm}_{\bbox{w}}(x;\sigma_0)\mapsto (-{\myrm w}^2)^{-\sigma_0}
\Phi^{(0)\mp}_{\bbox{w}}(x;\sigma_0).\]
Then a two-point function can be defined as
\[\label{twop-0-def}
\langle \stack{2}{x},\pm;\sigma_0|\stack{1}{x} ,\pm;\sigma_0 \rangle =
\int_{{\Bbb R}^3}\d^3 {{\bi w}}\, \Phi_{{\bi w}}^{(0)\pm}(\stack{1}{x};\sigma_0 )
\Phi_{{\bi w}}^{(0)\pm}(\stack{2}{x};\sigma_0^* )= \frac{1}{8}
{\cal W}^{(0)}(\stack{1}{x},\stack{2}{x}).\]
It is dS-invariant by the virtue of unitarity of the representation
$T^\pm_{\sigma_0}$:
\[{\cal W}^{(0)}
(\stack{1}{x}_{g},\stack{2}{x}_{g})={\cal W}^{(0)}(\stack{1}{x},\stack{2}{x})
\qquad  g\in SO(4,1).\]
We can obtain an another expression for
${\cal W}^{(0)}(\stack{1}{x},\stack{2}{x})$ passing to the integration over
3-sphere~\cite{24}. Let us consider the unit euclidean four-vector
$l_a$, $a,b=1,2,3,5$ dependent on the three-vector ${\bi w}$:
\[l^a_{{\bi w}}=\left( \mp \frac{2{{\bi w}}}{1+ w^2},
\frac{1-w^2}{1+ w^2}\right) \qquad
l^a_{{\bi w}}l^a_{{\bi w}}=1. \]
Then computing the Jacobian of the transformation from
${\bi w}$ to ${\bi l}_{{\bi w}}$ we obtain
\begin{equation}\label{twop-eucl}
{\cal W}^{(0)}(\stack{1}{x},\stack{2}{x})=\int_{S^3}\frac{\d^3{\bi l}}{l^5}
\left( \frac{\stack{1}{x}^0+l^a\stack{1}{x}^a}{R}
\right)^{-\i\mu R-3/2}
\left( \frac{\stack{2}{x}^0+l^a\stack{2}{x}^a}{R}
\right)^{\i\mu R-3/2}.
\end{equation}
The function ${\cal W}^{(0)}(\stack{1}{x},\stack{2}{x})$
coincides with the two-point function over the dS space considered
in~\cite{24}.  In general, the integral~(\ref{twop-eucl}) diverges because
$|\psi_0\rangle$ is not square-integrable over ${\Bbb R}^3$. We can make the
integral meaningful  passing to the complex
dS space with subsequent computation of the boundary values over the real dS
space~\cite{24}. Then the dS-invariance of two-point function remains
valid since the transformation rules~(\ref{transf}) remains correct.

Let us consider the domain ${\cal D}^\pm$ in the complex dS space
(we shall denote its points as $\zeta$) defined as
\[ \quad \pm\im \zeta^0 >0 \qquad
\im \zeta^A \im \zeta_A >0.\]
The domain ${\cal D}^+$ (${\cal D}^-$) is the domain of analyticity of
the functions
$\varphi^{(0)\pm}_{k}(\zeta;\sigma_0)$
($\varphi^{(0)\pm}_{k}(\zeta;\sigma_0^*)$). Then the integral
(\ref{twop-eucl}) converges at $\stack{1}{\zeta}\in {\cal D}^+$ and
$\stack{2}{\zeta}\in {\cal D}^-$ since the 3-sphere volume is finite.
Let us choose the points as
\begin{equation}\label{points-dS}
\stack{1}{\zeta}^A=(\i\cosh v,{\bi 0},i\sinh v) \qquad
\stack{2}{\zeta}^A=(-\i,{\bi 0},0) \qquad  v\in {\Bbb R}.
\end{equation}
Then using the formula~\cite{59}
\begin{equation}\label{BE123-8}
\mathop{_2 F_1} (a,b;c;z)=
\frac{2^{1-c}\Gamma(c)}{\Gamma(b)\Gamma(c-b)}
 \int_{0}^\pi \d\varphi
\, \frac{(\sin\varphi)^{2b-1}(1+\cos\varphi)^{c-2b}}{\left(
1-\frac{z}{2}+\frac{z}{2}\cos\varphi\right)^a}
\end{equation}
we obtain
\[\label{twop-cdS}
{\cal W}^{(0)}(\stack{1}{\zeta},\stack{2}{\zeta})=
\frac{\pi^2}{2} \e^{-\pi\mu R}\mathop{_2 F_1}\left( -\sigma_0^*,
-\sigma_0 ;2;\frac{1-\rho}{2}\right)\]
where $\rho=R^{-2} \stack{1}{\zeta}^A \stack{2}{\zeta}_A$.
The expression obtained in~\cite{24} is in fact equivalent to
the above expression to within a constant multiplier.

\subsection{Quantized spin zero field \label{field0-dS}}

Let us define the quantized spin zero field in the dS space as
\[\phi^{(0)}(x)=\int_{{\Bbb R}^3} \d^3 {{\bi w}}\, \left(
\Phi^{(0)+}_{{\bi w}}(x;\sigma_0 )a^{(+)} ({{\bi w}})+
\Phi^{(0)-}_{{\bi w}}(x;\sigma_0^* )a^{(-)\dagger}
({{\bi w}})\right)\]
where  $a^{(\pm)}({{\bi w}})$ and $a^{(\pm)\dagger} ({{\bi w}})$ are two sets of
bosonic creation-annihilation operators with the commutation relations
\[\label{comm-bosonic}
[a^{(\pm)}({{\bi w}}), a^{(\pm)\dagger} ({{\bi w}}')]=
\delta ({{\bi w}},{{\bi w}}') \]
where $\delta({\bbox{w}}_1 ,{\bbox{w}}_2)$ is the
$\delta$-function over ${\Bbb R}^3$ and all other commutators vanish.
Now  compute the propagator
\begin{equation}\label{3. 31a}
\Bigl[ \phi^{(0)}(\stack{1}{x}),\phi^{(0)\dagger}(\stack{2}{x})\Bigr] \equiv
\frac{1}{8}G^{(0)}(\stack{1}{x},\stack{2}{x})=
\frac{1}{8}\left( {\cal W}^{(0)}(\stack{1}{x},\stack{2}{x})-
{\cal W}^{(0)}(\stack{2}{x},\stack{1}{x}) \right)
\end{equation}
in the explicit form by passing to the complex dS space.
Considering the real dS space as a boundary of the domains
$(\stack{1}{\zeta},\stack{2}{\zeta})\in {\cal D}^\pm \times {\cal D}^\mp$,
let us denote the  boundary values of the function
${\cal W}^{(0)}(\stack{1}{\zeta},\stack{2}{\zeta})$
as ${\cal W}^{(0)\pm}(\stack{1}{x},\stack{2}{x})$. To compute these
boundary values we put
$\stack{1}{\zeta}=\stack{1}{x}+\i\stack{1}{\epsilon}$ and
$\stack{2}{\zeta}=\stack{2}{x}-\i\stack{2}{\epsilon},$
where $\stack{1}{\epsilon}$ and $\stack{2}{\epsilon}$ are two real
infinitesimal time-like forward four-vectors and then indeed
$(\stack{1}{\zeta},\stack{2}{\zeta})\in {\cal D}^+ \times {\cal D}^-$.
It is easily seen that
\[\stack{1}{\zeta}^A \stack{2}{\zeta}_A =\stack{1}{x}^A \stack{2}{x}_A +
\frac{\i}{\stack{1}{x}^5 \stack{2}{x}^5}(\stack{1}{\epsilon}+
\stack{2}{\epsilon})\cdot \left( \frac{\stack{2}{x}}{\stack{2}{x}^5}-
\frac{\stack{1}{x}}{\stack{1}{x}^5}\right).\]
Then under the above assumptions the sign of the imaginary part of
$\stack{1}{\zeta}^A \stack{2}{\zeta}_A$ does
not depend on the way in which
$\stack{1}{\epsilon}$ and $\stack{2}{\epsilon}$ tend to zero.
Let $\stack{2}{x}^\mu =0$ and $\stack{1}{x}\cdot \stack{1}{x}\geq 0$, then
\[\stack{1}{\zeta}^A \stack{2}{\zeta}_A =\stack{1}{x}^A \stack{2}{x}_A -
\i 0\varepsilon (\stack{1}{x}^0).\]
The case of backward
$\stack{1}{\epsilon}$ and $\stack{2}{\epsilon}$ (then
$(\stack{1}{\zeta},\stack{2}{\zeta})\in {\cal D}^- \times {\cal D}^+$)
may be considered in the completely analogous way. Then
\begin{equation}\label{pm-i0}
{\cal W}^{(0)\pm}(\stack{1}{x},\stack{2}{x})=
\frac{\pi^2}{2} \e^{-\pi\mu R}\mathop{_2 F_1}\left( -\sigma_0^*,
-\sigma_0 ;2;\frac{1-G \pm \i 0 \varepsilon(\stack{1}{x}^0)}{2}\right),
\end{equation}
where $G=R^{-2}\stack{1}{x}^A \stack{2}{x}_A$. As
$(\stack{1}{\zeta},\stack{2}{\zeta})\in {\cal D}^+ \times {\cal D}^-$ yields
$(\stack{2}{\zeta},\stack{1}{\zeta})\in {\cal D}^- \times {\cal D}^+$,
then we get
\begin{equation}\label{x1x2-x2x1}
{\cal W}^{(0)+}(\stack{1}{x},\stack{2}{x}) =
{\cal W}^{(0)-}(\stack{2}{x},\stack{1}{x})
\end{equation}
(cf. proposition 2.2 of~\cite{24}) and by the virtue of~(\ref{3. 31a})
the propagator is equal to
\[ G^{(0)}(\stack{1}{x},\stack{2}{x})=
{\cal W}^{(0)+}(\stack{1}{x},\stack{2}{x})-
{\cal W}^{(0)-}(\stack{1}{x},\stack{2}{x}).\]
The function
${\cal W}^{(0)-}(\stack{1}{x},\stack{2}{x})$ coincides to within the constant
multiplier with the propagator obtained in~\cite{76} starting
from the demands of satisfaction of
the Klein-Gordon equation and the boundary conditions. This function may be
obtained by the summation over the modes~\cite{33} and by the method of
discrete lattice~\cite{85} too.

As we assume that $\stack{1}{x}\cdot \stack{1}{x}\geq 0$ then
$\frac{1-G}{2}\geq 1$, but the integral~(\ref{BE123-8}) with which we define
the hypergeometric function, makes no sense at
$z\in [1,+\infty)$ and then  demands the analytic continuation
in the domain which contains the mentioned interval. To this end we shall use
the formulas~\cite{59}
\begin{eqnarray}
u_1 =\frac{\Gamma (c)\Gamma(b-a)}{\Gamma(c-a)\Gamma(b)}u_3 +
\frac{\Gamma(c)\Gamma(a-b)}{\Gamma(c-b)\Gamma(a)} u_4 \label{u1}\nonumber \\
u_2 =\frac{\Gamma (a+b+1-c)\Gamma(b-a)}{\Gamma(b+1-c)\Gamma(b)}
\e^{\mp \i\pi a}u_3 +
\frac{\Gamma(a+b+1-c)\Gamma(a-b)}{\Gamma(a+1-c)\Gamma(a)}
\e^{\mp \i\pi b} u_4  \nonumber
\end{eqnarray}
where the upper or lower sign should be chosen depending on the $\im z$
greater or smaller  than zero, and $u_1,\ldots,u_4$ are the Kummer solutions
of the hypergeometric equation:
\begin{eqnarray}
u_1 =\mathop{_2 F_1}(a,b;c;z)\nonumber \\
u_2 =\mathop{_2 F_1}(a,b;a+b+1-c;1-z)\nonumber \\
u_3 =(-z)^{-a}\mathop{_2 F_1}(a,a+1-c;a+1-b;z^{-1})\nonumber \\
u_4 =(-z)^{-b}\mathop{_2 F_1}(b,b+1-c;b+1-a;z^{-1}) .\nonumber
\end{eqnarray}
The functions $u_1 ,u_3, u_4$ are holomorphic at $z<0$. Then
at $a+b+1=2c$
\begin{equation}\label{app1}
\left. u_2 \right|_{z-\i 0}^{z+\i 0} =\i (\e^{\pi\mu R}+\e^{-\pi\mu R})
\theta (-z)u_1 \qquad  z\not= 0.
\end{equation}
To obtain the behavior of $u_2$ at
$z=0$ we shall use the formula~\cite{59}
\begin{eqnarray}
\fl \mathop{_2 F_1}(a,b;a+b-m;z)\frac{1}{\Gamma (a+b-m)}\nonumber \\
\lo= \frac{\Gamma(m) (1-z)^{-m}}{\Gamma(a)\Gamma(b)}
\sum_{n=0}^{m-1}\frac{(a-m)_n (b-m)_n}{(1-m)_n n!}(1-z)^n
\nonumber \\
+ \frac{(-1)^m}{\Gamma(a-m)\Gamma(b-m)}\sum_{n=0}^{\infty}
\frac{(a)_n (b)_n}{(n+m)_n n!}[\overline{h}_n -\ln (1-z)](1-z)^n \nonumber
\end{eqnarray}
which holds at $|\arg (1-z)|<\pi$, and the formulas
\begin{eqnarray}
z^{-1}|_{z-\i 0}^{z+\i 0}=-2\pi \i\delta(z) \nonumber \\
\Gamma \left(\frac{1}{2}+z\right)\Gamma \left(\frac{1}{2}-z\right)
=\frac{\pi}{\cos (\pi z)}. \nonumber
\end{eqnarray}
Then using~(\ref{app1}) we obtain
\begin{equation}\label{F|+-}
\eqalign{
\fl\left.
\mathop{_2 F_1}(-\sigma_0 ,-\sigma_0^* ;2;z)\right|_{z-\i 0}^{z+\i 0}=
-\i (\e^{\pi\mu R}+\e^{-\pi\mu R}) \\
\times \left( \frac{\delta (1-z)}{\mu^2 +(4R^2)^{-1}}-\theta(z-1)
\mathop{_2 F_1}(-\sigma_0 ,-\sigma_0^* ;2;1-z)\right).
}
\end{equation}
Putting together the above expression and~(\ref{pm-i0})  we finally obtain
\begin{equation}\label{G0-dS}
\eqalign{
\fl G^{(0)}(\stack{1}{x},\stack{2}{x})=-\i\pi^2 (1+\e^{-2\pi\mu R})
\varepsilon (\stack{1}{x}^0 -\stack{2}{x}^0)  \\
\times\left[ \frac{\delta (1+G)}{\mu^2 +(4R^2)^{-1}}-\frac{1}{2}
\theta \left(-\frac{1+G}{2}\right) \mathop{_2 F_1}
\left(-\sigma_0 ,-\sigma_0^* ;2; \frac{1+G}{2}\right)\right].
}
\end{equation}
To within the constant multiplier, the above expression coincides with that
obtained previously starting from the demands of satisfaction of the
Klein-Gordon equation and the boundary conditions~\cite{77/81/82}.

\section{Spinor field}

\subsection{The Dirac equation}

Introducing the matrices
\[\gamma ^{5}={\i}\gamma ^{0}\gamma ^{1}\gamma ^{2}\gamma ^{3} \qquad
\tgam^{\mu} =-{\i}\gamma ^{5}\gamma ^{\mu} \qquad
 \tgam^{5}=\i\gamma ^{5} \]
we can write down the generators of the four-spinor representation of the
dS group in the five-dimensional form:
\begin{equation}\label{5. 39a}
J^{(s)AB}=\frac{1}{4} [\tgam ^{A}, \tgam ^{B}].
\end{equation}
The equalities
\begin{equation}\label{5.  40a}
 \tgam^{A}\tgam^{B}+
\tgam^{B}\tgam^{A}=2\eta ^{AB}
\end{equation}
\begin{equation}
\label{5.  40b}
\tgam^{A}\tgam^{B}\tgam^{C}=
\eta ^{AB}\tgam^{C}+\eta ^{BC}\tgam^{A}-
\eta ^{AC}\tgam^{B}+\frac{1}{2}
\varepsilon ^{ABCDE}\tgam_{D}\tgam_{E}
\end{equation}
\begin{equation}
\label{5.  40c}
\eqalign{
\tgam^{A}\tgam^{B}\tgam^{C}\tgam^{D}
=\eta ^{AB}\tgam^{C}\tgam^{D}+
\eta ^{BC}\tgam^{A}\tgam^{D}-
\eta ^{AC}\tgam^{B}\tgam^{D}+ \\
+2(\eta ^{AD}J^{(s)BC}+\eta ^{CD}J^{(s)AB}-\eta ^{BD}J^{(s)AC} )-
\varepsilon ^{ABCDE}\tgam_{E}
}
\end{equation}
hold. With the help of the above expressions
and~(\ref{5.  26}),(\ref{5.  28}) we obtain
\begin{equation}\label{5.  41}
 R^{2}C^{(s)}_{2}=5/2  \qquad
\label{5.  42}
 W^{(s)}_{A}=\frac{3}{4}\tgam_{A} \qquad
\label{5.  43}
  R^{2}C^{(s)}_{4}=\frac{45}{16}.
  \end{equation}
Comparing the above expression with~(\ref{5.  29})
we see that it is the representation $\bpi_{3/2, 3/2}$.
We shall choose the standard form of $\gamma$-matrices. Then it is easy to
show that the explicit form of generators is
\begin{eqnarray}\label{3.  11a}
\bPi^{+(s)}=\frac{1}{R}\left(
\begin{array}{rr}
0 & -\bsigma \\
0 & 0
\end{array}
\right) \qquad
\bPi^{-(s)}=\frac{1}{R}\left(
\begin{array}{rr}
0 & 0 \\
\bsigma & 0
\end{array}
\right) \\
\label{3.  11c}
P^{0(s)}=\frac{1}{2}\left(
\begin{array}{rr}
1 & 0 \\
0 & -1
\end{array}
\right) \qquad
J_{ik}^{(s)}=-\i\varepsilon_{ikl}
\left(
\begin{array}{ll}
\sigma^{l} & 0 \\
0 & \sigma^{l}
\end{array}
\right) . \nonumber
\end{eqnarray}
We denote matrices of finite transformations  as $U(g)$. Then we obtain
\begin{eqnarray}\label{3.  13a}
 U(\Theta_{\pm}({{\bi a}}))=1-\bPi^{\pm (s)}{{\bi a}}R \nonumber \\
\label{3.  13b}
U(\Theta_{0}(\varepsilon))=\exp (-P^{(s)}_{0}\varepsilon). \nonumber
\end{eqnarray}

Let us consider the representation
$\bpi_{3/2, 3/2}\otimes \bnu_{m, 0}$.
Its generators are the sum of generators~(\ref{5. 34}) ({\it orbital}
part) and generators~(\ref{5. 39a}) ({\it spin} part).
Then the second-order Casimir operator is equal to
\[ C_{2}=C^{(l)}_{2}+C^{(s)}_{2}-R^{-2}J^{(s)AB}J^{(l)}_{AB}.  \]
Denoting
$\hat{\nabla}_{\rm dS}=-R^{-1}J^{(s)AB}J^{(l)}_{AB}$
we obtain
\begin{equation}\label{5.  45}
C_{2}=\Box  +\frac{\hat{\nabla}_{\rm dS}}{R}+\frac{5}{2R^{2}}.
\end{equation}
To compute the fourth-order Casimir operator we write according
to~(\ref{5.  35}) and~\rf{5.  42}:
\[ RW_{A}=
-\frac{1}{8}\varepsilon _{ABCDE}\tgam^{B} \tgam^{C}
J^{(l)DE}+\frac{3}{4}\tgam_{A}.  \]
For squaring $W_{A}$  it is necessary to use
formulas~\rf{5.  40a}-(\ref{5. 40c}). The result obtained
\[ C_{4}=\frac{3}{4}\Box+\frac{3\hat{\nabla}_{\rm
dS}}{4R}+\frac{45}{16R^{2}}\]
is in agreement with~(\ref{5.  32})  at $s=1/2$ and~\rf{5.  45}.
From the second Shur's lemma  follows that the operators
$\hat{\nabla}_{\rm dS}$  and $\Box$
should have fixed eigenvalues in the irreducible representations.
Then using~(\ref{5.  31}),~\rf{5.  45} and the equality
\[\hat{\nabla}_{\rm dS}^{2}=\frac{1}{4R^{2}}\tgam^{A}\tgam^{B}\tgam^{C}
\tgam^{D}J^{(l)}_{AB}J^{(l)}_{CD}=\Box -3\hat{\nabla}_{\rm dS}/R\]
we obtain the quadratic equation for eigenvalues of
$\hat{\nabla}_{\rm dS}$. Solving it yields
\begin{eqnarray}\label{5. 46a}
\hat{\nabla}_{\rm dS}=-2R^{-1}\pm {\i}\mu \\
\Box =-\mu ^{2}\mp \i R^{-1}\mu -2R^{-2} \nonumber
\end{eqnarray}
where $\mu ^{2}=m^{2}-\frac{1}{4R^{2}}.$ As
$m^{2}>1/4R^{2}$ (see subsection 3.1),
then $\mu$ is a real number. The appearance of two signs
indicates that two identical irreducible representations have appeared:
\[\bnu_{m, 0} \otimes \bpi_{3/2, 3/2}=2\bnu_{m, 1/2}.\]
Using~(\ref{5.  34})  we can write
\[\hat{\nabla}_{\rm dS}=\Gamma ^{\mu}\partial _{\mu} \qquad
 \Gamma ^{\mu}=\chi  \gamma ^{\mu}
+\frac{1}{2R} [\gamma ^{\mu}, \gamma _{\nu}]x^{\nu}.\]
Choosing the representation which corresponds to the lower sign
in~(\ref{5. 46a}) we finally obtain the group theoretical Dirac
equation over the dS space:
\begin{equation}\label{5.  47}
{\i}\Gamma ^{\mu}\partial _{\mu}\psi-(\mu -2\i R^{-1})\psi =0.
\end{equation}
This was well known previously out of the context of dS group irreducible
representations~\cite{Gursey}.

The above equation
 admits the transformation into the
covariant form. To this end let us perform the transformation
$\Psi =V\psi,$ where
\[\label{5.  47a}
V=(1-\varepsilon_{\mu}\varepsilon^{\mu} )^{-1/2}
(1+\gamma_{\mu}\varepsilon^{\mu} ) \qquad
\varepsilon ^{\mu}=\frac{x^{\mu}}{R(\chi +1)}.\]
Then~\rf{5. 47} passes into
 \begin{equation}\label{5.  48}
 \i V\Gamma ^{\mu}V^{-1}(\partial _{\mu}\Psi
 +(V\partial _{\mu}V^{-1})\Psi )-(\mu -2\i R^{-1})\Psi =0.
\end{equation}
It is easy to show that
\begin{equation}\label{5.  49}
\eqalign{
 V\Gamma ^{\mu}V^{-1}=e^{\mu}_{(\nu )}\gamma ^{\nu} \\
\label{5.  50}
  \partial _{\mu}+V\partial _{\mu}V^{-1} ={\cal D}_{\mu}-
\frac{1}{2R}\gamma _{\mu}
}
\end{equation}
where $e^{(\mu)}_{\nu}$ is the vierbein which is orthonormal with respect to
the metric~(\ref{5. 21}):
 \[\label{5.  51}
e^{(\mu) \nu}=\eta ^{\mu \nu}+\frac{x^{\mu}x^{\nu}}{R^{2}(\chi +1)}\]
and  ${\cal D}_{\mu}$ is the spinor covariant derivative
\[\fl {\cal D}_{(\mu)}=e_{(\mu)}^\nu
\partial_{\nu}-\frac{1}{2}J^{(s)\nu\rho} G_{\nu\rho\mu} \qquad
G_{\nu\rho\mu}=
e_{(\nu);\kappa}^\sigma e_{(\rho)\sigma}e^\kappa_{(\mu)}
=\frac{1}{R^{2}(\chi +1)}(x_{\nu}\eta_{\mu \rho}-
x_{\rho}\eta_{\nu\mu}).  \]
Then putting together~(\ref{5. 48})-(\ref{5. 50}) we finally obtain
\[\i\gamma ^{\mu} e^{\nu}_{(\mu )} {\cal D}_{\nu}\Psi =\mu \Psi. \]
Another more complicated ways of transformation of (Anti-)dS-invariant
Dirac equation to the covariant one were proposed in~\cite{61,gath}.

\subsection{Spinor coherent states}
In~\cite{dS-PLB} was shown that the solutions of equation~(\ref{5. 47}) may
be obtained in the form of the product of scalar CS with an imaginary shift
of a mass, and basic Dirac 4-spinors. However, there is not the natural
action of the full dS group over these 4-spinors.
This difficulty may be overcome if we consider
$4\times 2$-matrices whose columns are  the mentioned 4-spinors. Indeed,
let us denote the constant $4\times 2$-matrices as
$A,A',A''$ and  define over such  matrices the weak equivalence
relation $\sim$ and the strong one $\simeq$ as
\begin{eqnarray}
A'\sim A'' \Leftrightarrow A'= A''B \qquad  B\in GL(2,{\Bbb C})
\nonumber \\
A'\simeq A'' \Leftrightarrow A'= A''B \qquad  B\in SU(2). \label{3. 35}
\nonumber
\end{eqnarray}
Also define the product of two $4\times 2$-matrices
$A'$ and $A''$ as $A'\overline{A''}$, where the upper line
denotes the Dirac conjugation. Consider the left action of four-spinor
representation of the dS group over these matrices:
 $g:\ A\mapsto U(g)A.$ It is easy to show that the matrices
\[\label{vectors+-}
|+\rangle=\left(
\begin{array}{l}
I_2 \\ 0_2
\end{array}
\right)   \qquad
|-\rangle=\left(
\begin{array}{l}
0_2 \\ I_2
\end{array}
\right)\]
(where $I_2$ is the unit $2\times 2$-matrix) are invariant
under transformations which belong to the subgroups
${\cal K}^\pm \equiv {\cal T}^\pm \circledS ({\cal T}^0 \otimes {\cal R})$
to within the weak equivalence relation.
In the terms of the strong equivalence relation we have
\[ U(h)|\pm\rangle \simeq (\alpha^\pm_{\bbox{v}}(h))^{-1/2}
|_{\bbox{v}=\bbox{0}}|\pm\rangle \qquad  h\in {\cal K}^\pm .\]
From the other hand, it is easily seen that the subgroups
${\cal K}^\pm$ are the stability subgroups of the vector
$\bbox{w}=\bbox{0}$ concerning the conformal action~(\ref{3. 15})
of the dS group. This allows us to identify the
$SO(4,1)/{\cal K}^\pm$ space with the space ${\Bbb R}^3$ of vectors
$\bbox{w}$. As the lifting from the
$SO(4,1)/{\cal K}^\pm$ space to the dS group we shall take the transformation
which transforms the origin into the point ${\bi w}$:
\[ SO(4,1)/{\cal K}^\pm \ni \bbox{w}\mapsto
g_{\bbox{w}}=\Theta_\mp (-\bbox{w}) \in SO(4,1).\]
Then the CS system for the $SO(4,1)/{\cal K}^\pm$ space
being dS-invariant to within the weak equivalence relation is
\begin{eqnarray}
|\bbox{w}\pm\rangle =U(g_{\bbox{w}})|\pm\rangle \nonumber \\
|\bbox{w}+\rangle=\left(
\begin{array}{l}
I_2 \\ \bsigma\bbox{w}
\end{array}
\right)   \qquad
|\bbox{w}-\rangle=\left(
\begin{array}{l}
-\bsigma\bbox{w} \\ I_2
\end{array}
\right). \nonumber
\end{eqnarray}
With the help of~(\ref{foliation}) the transformation properties of
these vectors  with respect to the strong equivalence relation
may be written as
\begin{equation}\label{transf-dS-1/2}
 U(g_1 )|\bbox{w}\pm\rangle \simeq
(\alpha^\pm_{\bbox{v}}(g_{\bbox{w}'}^{-1}g_1 g_{\bbox{w}})
)^{-1/2}|_{\bbox{v}=\bbox{0}}
|{\bbox{w}}_{g_1} \pm\rangle \qquad  g_1 \in {\cal G}.
\end{equation}
As the transformations $T^\pm_\sigma (g)$ compose a representation of the dS
group then
\[\alpha^\pm_{\bbox{v}}(g_2 g_1 )=\alpha^\pm_{\bbox{v}}(g_2 )
\alpha^\pm_{\bbox{v}'}(g_1) \qquad  g_1 ,g_2 \in {\cal G}
\qquad  \bbox{v}'=\bbox{v}_{g_2^{-1}}.\]
Then using the above expression and~(\ref{foliation}) we get
\[\alpha^\pm_{\bbox{v}}(g_1)=\alpha^\pm_{\bbox{v}'}
(g_{\bbox{w}'}^{-1}g_1 g_{\bbox{w}})  \qquad
\bbox{v}'=\bbox{v}_{g_{\bbox{w}'}^{-1}} \qquad
\bbox{w}'=\bbox{w}_{g_1} .\]
Putting $\bbox{v}=\bbox{w}'$ in the above expression, we can rewrite the
transformation properties~(\ref{transf-dS-1/2}) as
\begin{equation}\label{3. 37}
(\alpha^\pm_{\bbox{w}}(g))^{1/2}U(g)|\bbox{w}_{g^{-1}}\pm\rangle
\simeq |\bbox{w}\pm\rangle \qquad  g\in {\cal G}.
\end{equation}
It is easy to show that the equalities
\begin{eqnarray}\label{3. 36}
\left(\gamma\cdot k_{{\bi w}}\mp 1 \right)|{{\bi w}}\pm\rangle =0 \\
|{\bi w}\pm\rangle\langle {\bi w}\pm | =\frac{1- w^2}{2}
(\gamma\cdot k_{{\bi w}}\pm 1) \label{<w|w>}
\end{eqnarray}
are correct. Now let us construct  the $4\times 2$-matrix functions
\[\Phi_{{\bi w}}^{(1/2)\pm}(x)=
\Phi_{{\bi w}}^{(0)\pm}(x;\sigma_0 -1/2)|{{\bi w}}\pm\rangle.\]
Using~(\ref{3. 36}) we obtain that they obey~(\ref{5. 47}):
\[ (\i\hat{\nabla}_{\rm dS} -\mu+2\i R^{-1})
\Phi_{\bbox{w}}^{(1/2)\pm}(x)=0.\]
These solutions are much simpler that
those obtained by the method of separation of variables~\cite{16/65,87}.

From the transformation properties~(\ref{transf})
and~(\ref{3. 37}) it follows that under the transformations from the dS group
the functions $\Phi_{{\bi w}}^{(1/2)\pm}(x)$ transform just as the functions
$\Phi_{{\bi w}}^{(0)\pm}(x;\sigma_0)$, to within the constant matrix
transformation:
\begin{equation}\label{3. 27}
\Phi_{{\bi w}}^{(1/2)\pm}(x_{g})\simeq
(\alpha^{\pm}_{{\bi w}}(g))^{\sigma_0} U(g)
\Phi_{{{\bi w}}'}^{(1/2)\pm}(x)
\end{equation}
where ${{\bi w}}'={{\bi w}}_{g^{-1}}$.
As the inversion $\bbox{w}\mapsto -\bbox{w}/{\myrm w}^2$ yields
\[\Phi^{(1/2)\pm}_{-\bbox{w}/{\myrm w}^2}(x) \simeq
-i (-{\myrm w}^2)^{-\sigma_0}\Phi^{(1/2)\mp}_{\bbox{w}}(x)\]
then the functions $\Phi_{{\bi w}}^{(1/2)+}(x)$ and
$\Phi_{{\bi w}}^{(1/2)-}(x)$ yield the
same two-point function. Let us define it as follows:
\[ \fl \frac{1}{8}{\cal W}^{(1/2)} (\ksi{1},\ksi{2})=
\int_{{\Bbb R}^3}\d^3 {{\bi w}}\,\Phi^{(1/2)+}_{{\bi w}}(\ksi{1})
\overline{\Phi}\vphantom{\Phi}^{(1/2)+}_{{\bi w}}(\ksi{2})
= \int_{{\Bbb R}^3}\d^3 {{\bi w}}\,\Phi^{(1/2)-}_{{\bi w}}(\ksi{1})
\overline{\Phi}\vphantom{\Phi}^{(1/2)-}_{{\bi w}}(\ksi{2}). \]
From~(\ref{3. 27})  follows that it is dS-invariant in the
sense that at $g\in\cal G$
\[\label{3.  33}
{\cal W}^{(1/2)}(\ksi{1}_{g},\ksi{2}_{g})=
U(g){\cal W}^{(1/2)}(\ksi{1},\ksi{2})\overline{U}(g).\]
Using~(\ref{<w|w>})  it is easy to show that
\[ \fl {\cal W}^{(1/2)}(\ksi{1},\ksi{2})
= \frac{1}{2}\int_{S^3}\frac{\d^3{\bi l}}{l^5}
\left( \frac{\stack{1}{x}^0+l^a\stack{1}{x}^a}{R}
\right)^{-\i\mu R-2}
\left( \frac{\stack{2}{x}^0+l^a \stack{2}{x}^a}{R}
\right)^{\i\mu R-2}
(\gamma^0 +\bgamma {\bi l}+l^5 ). \label{dirac-3-sphere}\]
As the functions $\Phi_{\bi w}^{(1/2)\pm}(\zeta)$ inherit the
analyticity properties of functions $\varphi_{k}^{(0)\pm}(\zeta;\sigma_0)$
over the complex dS space, then, similarly to in the spin zero case, the
function ${\cal W}^{(1/2)}(\stack{1}{x},\stack{2}{x})$ converges at
$(\stack{1}{\zeta},\stack{2}{\zeta})\in {\cal D}^+ \times {\cal D}^- .$
Choosing the points according
to~(\ref{points-dS}) and using the equality
\begin{eqnarray}
\fl\int_{0}^\pi \d\theta \, \sin^2 \theta \cos\theta
(\cosh v+\sinh v\cos\theta )^{-\i\mu R-2}
=\frac{\pi}{\sinh v} \nonumber\\
\fl\times\left(
\mathop{_2 F_1}
\left( 1-\frac{\i\mu R}{2}, \frac{1+\i\mu R}{2};2; -\sinh^2 v\right)
-\cosh v \mathop{_2 F_1}
\left( 1+\frac{\i\mu R}{2}, \frac{1-\i\mu R}{2};2; -\sinh^2 v\right)
\right) \nonumber
\end{eqnarray}
we obtain
\begin{equation}\label{twop-1/2-cdS}
\eqalign{
\fl {\cal W}^{(1/2)}(\stack{1}{\zeta},\stack{2}{\zeta})
 = \frac{\pi^2\e^{-\pi\mu R}}{\mu -\i R^{-1}} \\
\times\tilde{\gamma}\vphantom{\gamma}_A
\stack{1}{\zeta}^A \left( \i\hat{\nabla}_{\rm dS}-\mu  +\i R^{-1}\right)
\mathop{_2 F_1}\left(2-\i\mu R,1+\i\mu R;2;\frac{1-\rho}{2}\right)\gamma^5
}
\end{equation}
where the operator $\hat{\nabla}_{\rm dS}$  acts onto the coordinates
$\stack{1}{\zeta}$.

\subsection{Quantized spinor field}

To construct a quantized spinor field, let us  use  the equality
\[ R^{-1}\tgam^A x_A
\left(\i\hat{\nabla}_{\rm dS} -\mu + 2\i R^{-1}\right) R^{-1} \tgam^B
x_B= \i\hat{\nabla}_{\rm dS} +\mu +2\i R^{-1}.\]
From here follows that
 if the function $\psi$ obey the Dirac equation~(\ref{5. 47}),
then the function $\tgam^A x_A \psi$ obey the same equation with the opposite
sign of $\mu$. Then the functions
\[\tilde{\Phi}\vphantom{\Phi}_{{\bi w}}^{(1/2)\pm}(x)=R^{-1}\tgam^A x_A
\Phi_{{\bi w}}^{(0)\pm}(x;\sigma_0^* -1/2)|{{\bi w}}\pm\rangle\]
obey equation~(\ref{5. 47}). Let us introduce also two sets of fermionic
creation-annihilation operators $b^{(\pm)}({{\bi w}})$
and $b^{(\pm)\dagger}({{\bi w}})$, which at the same time
 are  the matrices of dimensionality $2\times 1$ and $1\times 2$,
respectively, and obey the anticommutation relations
\begin{equation}\label{comm-fermionic}
\{ b^{(\pm)}({{\bi w}}) , b^{(\pm)\dagger}({{\bi w}}')\}
=\delta ({{\bi w}},{{\bi w}}')
\left(
\begin{array}{ll}
1 & 0 \\
0 & 1
\end{array}
\right)
\end{equation}
and all other anticommutators vanish.
Then we can construct the quantized spinor field as
\begin{equation}\label{field-dS-1/2}
\phi^{(1/2)}(x)=\int_{{\Bbb R}^3} \d^3 {{\bi w}}\,\left( \Phi^{(1/2)+}_{{\bi
w}} (x)b^{(+)}({{\bi w}}) + \tilde{\Phi}\vphantom{\Phi}_{{\bi w}}^{(1/2)-}(x)
b^{(-)}({{\bi w}}) \right).
\end{equation}
Using~(\ref{twop-1/2-cdS}) it is easy to show that the two-point function
which corresponds to the solutions
$\tilde{\Phi}\vphantom{\Phi}_{{\bi w}}^{(1/2)\pm}(x)$ is equal to
\begin{equation}\label{z2z1-z1z2}
\eqalign{
\int_{{\Bbb R}^3}\d^3 {{\bi w}}\,
\tilde{\Phi}\vphantom{\Phi}_{{\bi w}}^{(1/2)\pm}(\stack{1}{\zeta})
\overline{\tilde{\Phi}}
\vphantom{\Phi}_{{\bi w}}^{(1/2)\pm}(\stack{2}{\zeta})   \\
\lo= R^{-2} \tgam^A \stack{1}{\zeta}_A
{\cal W}^{(1/2)} (\stack{2}{\zeta},\stack{1}{\zeta})
(\gamma^5 \tgam^B \stack{2}{\zeta}_B \gamma^5)=
-{\cal W}^{(1/2)} (\stack{1}{\zeta},\stack{2}{\zeta}).
}
\end{equation}
Further, the hypergeometric functions in the r.h.s.
of~(\ref{twop-1/2-cdS}) differ from
${\cal W}^{(0)} (\stack{1}{\zeta},\stack{2}{\zeta})$ only by the constant
multiplier and the imaginary shift of mass: $\mu\rightarrow\mu+iR^{-1}$.
Then computing the
difference of its values on the edges of the cut
$z\in [1,+\infty)$ we can use~(\ref{F|+-}). Passing to
the boundary values equation (\ref{z2z1-z1z2}) yields
\[ R^{-2} \tgam^A \stack{1}{x}_A{\cal W}^{(1/2)+}
 (\stack{2}{x},\stack{1}{x}) (\gamma^5 \tgam^B \stack{2}{x}_B \gamma^5)=
-{\cal W}^{(1/2)-} (\stack{1}{x},\stack{2}{x}) \]
which is analogous to Equation~(\ref{x1x2-x2x1}) for the spin zero case.
Then using~(\ref{F|+-}) for the spin 1/2 propagator
\[\{ \phi^{(1/2)}_\alpha
(\ksi{1}),\overline{\phi}\vphantom{\phi}^{(1/2)}_\beta (\ksi{2}) \} \equiv
\frac{1}{8} G^{(1/2)}_{\alpha\beta}(\ksi{1},\ksi{2})
=\frac{1}{8} \left( {\cal W}^{(1/2)+}_{\alpha\beta} (\ksi{1},\ksi{2})-
{\cal W}^{(1/2)-}_{\alpha\beta}(\ksi{1},\ksi{2})\right) \]
where $\alpha,\beta=1,\ldots,4$ are spinor indices, we finally obtain
\begin{eqnarray}
\fl G^{(1/2)}(\ksi{1},\ksi{2})
=\frac{2\pi^2}{\mu -\i R^{-1}}(1-\e^{-2\pi\mu R})
\varepsilon (\stack{1}{x}^0 -\stack{2}{x}^0)
\tgam_A \stack{1}{x}^A \left( \i\hat{\nabla}_{\rm dS}-\mu+\i R^{-1}\right)
 \nonumber \\ \times
\left[ \frac{\delta(1+G)}{\mu(\mu+\i R^{-1})}-
\frac{1}{2}\theta \left(-\frac{1+G}{2}\right) \mathop{_2 F_1}
\left(2-\i\mu R ,1+\i\mu R ;2; \frac{1+G}{2}\right)\right]\gamma^5 ,
 \nonumber
\end{eqnarray}
where the operator
$\hat{\nabla}_{\rm dS}$ acts onto the coordinates $\stack{1}{x}$. The above
expression coincides with the solution of Cauchy problem for the Dirac
equation over the dS space obtained in~\cite{61}, to within a constant
multiplier.

\section{Concluding remarks}

To summarize the results of this  paper,
we can say that the CS method allow us to quantize massive spin~0 and~1/2
fields over the dS space in the uniform way.
Both in the spin zero case and in the spin~1/2 one
the starting-point is the invariant wave equations which
correspond to  irreducible representations of the dS group. The solutions
of these equations are constructed using  CS  for the dS group; in the spin
zero case the dS-invariant Klein-Gordon equation is satisfied by the scalar
CS itself. In the spin~1/2 case the solutions of dS-invariant Dirac equation
are constructed from two different CS systems which correspond to
different representations of the dS group and different stationary subgroups.
Both in the spin zero case and in the spin~1/2 one these sets of solutions
possess the same transformation properties under the dS group, with
only difference that the constant matrix transformation is added in the
spin~1/2 case.

From these sets of solutions we can construct the two-point functions
${\cal W}^{(s)}(\stack{1}{x},\stack{2}{x})$ which have the following
properties:

\begin{enumerate}
\item dS-invariance:
\[ {\cal W}^{(1/2)}(\ksi{1}_{g},\ksi{2}_{g})=
U_s (g){\cal W}^{(1/2)}(\ksi{1},\ksi{2})\overline{U}_s (g)\]
where $U_s (g)$ is the identical representation at $s=0$ and the four-spinor
representation at $s=1/2$.

\item Causality:
\[ {\cal W}^{(s)}(\stack{1}{x},\stack{2}{x})=
{\cal W}^{(s)}(\stack{2}{x},\stack{1}{x}) \qquad
\stack{1}{x}_A \stack{2}{x}^A >-R^2.\]

\item Regularized function ${\cal W}^{(s)}(\stack{1}{x},\stack{2}{x})$
is the boundary value of the function
${\cal W}^{(s)}(\stack{1}{\zeta},\stack{2}{\zeta})$
which is analytic in  certain domain of the complex dS space.
\end{enumerate}
For the spin zero case the above properties were proved in~\cite{24}; but
also in this case the CS method gives the sufficient simplification
since the property~1 is found almost obvious. 
Defining the creation-annihilation operators so that they
possess the necessary (anti)commutation relations, we can construct the
 quantized fields $\phi^{(s)}(x)$ using the mentioned sets of solutions;
the propagators of these fields are equal to
\[ [\phi^{(s)}(\stack{1}{x}),
\overline{\phi}\vphantom{\phi}^{(s)}(\stack{2}{x})
]_\pm =\frac{1}{8} \left( {\cal W}^{(s)}(\stack{1}{x},\stack{2}{x})-
{\cal W}^{(s)}(\stack{2}{x},\stack{1}{x}) \right) \]
and therefore are dS-invariant and causal automatically.

\ack

I am grateful to Yu P Stepanovsky for the constant support during
the work and to W Drechsler and Ph Spindel sending me  copies of their
papers~\cite{coher9/78,caus4/5,80}.

\end{document}